\documentclass{eptcs}

\usepackage{graphicx}
\usepackage{comment}
\usepackage{listings}
\usepackage{cite}
\newcommand{\bigcell}[2]{\begin{tabular}{@{}#1@{}}#2\end{tabular}}
\newcommand{\prob}{{\sc ProB}}
\newcommand{\pyb}{{\sc pyB}}
\title{Checking Computations of Formal Method Tools - A Secondary Toolchain for \prob}
\author{
	John Witulski \qquad\qquad  Michael Leuschel
\institute{
 Institut f\"{u}r Informatik, Heinrich-Heine Universit\"{a}t D\"{u}sseldorf\\
  Universit\"{a}tsstr. 1, D-40225 D\"{u}sseldorf\\
  \email{\{witulski, leuschel\}@cs.uni-duesseldorf.de}}
}

\date{\today}
\begin{document}

\pagestyle{empty}

\def\titlerunning{A Secondary Toolchain for \prob}
\def\authorrunning{Leuschel, Witulski}
\maketitle

\begin{abstract}
We present the implementation of \pyb, a predicate- and expression-checker for the B language.
The tool is to be used for a secondary tool chain for data validation and data generation, with \prob\ being used in the primary tool chain. 
Indeed, \pyb\ is an independent cleanroom-implementation which
 is used to double-check solutions generated by \prob, an animator and model-checker for B specifications.
One of the major goals is to use \prob\ together with \pyb\ to generate reliable outputs for high-integrity safety critical applications.
Although \pyb\ is still work in progress, the ProB/pyB toolchain has already been successfully tested on various industrial B machines and data validation tasks.
\end{abstract}

\section{Introduction}
\subsection{Motivation}

The success of formal methods in practice depends on fast, scalable but also reliable tools.
Indeed, a bug inside a tool can have disastrous 
consequences in the context of safety critical software.

One solution to this problem is to prove the correctness of the tool itself.
However, many tools used in the context of formal methods 
 consist of tens or hundreds of thousands of lines of code which haver evolved over long periods of time.
Reimplementing these tools to be correct by construction or verifying these tools formally a posteriori is often impractical.

An alternate solution to increase the trust in the output of the tool by using a {\em double chain}, i.e. validating the output of the main tool by a second, 
independently developed  tool.
In some cases, the second tool can also be much simpler, as its purpose is just checking an output,
not producing it in the first place.
This is the solution we have pursued in this paper for the \prob\ tool \cite{LeuschelButler:FME03,LeuschelPlagge:STTT10},
 which is an animator, model checker and constraint solver for the B-method \cite{Abrial:BBook}.

\begin{figure}[ht]
	\centering
  	\includegraphics[width=260px]{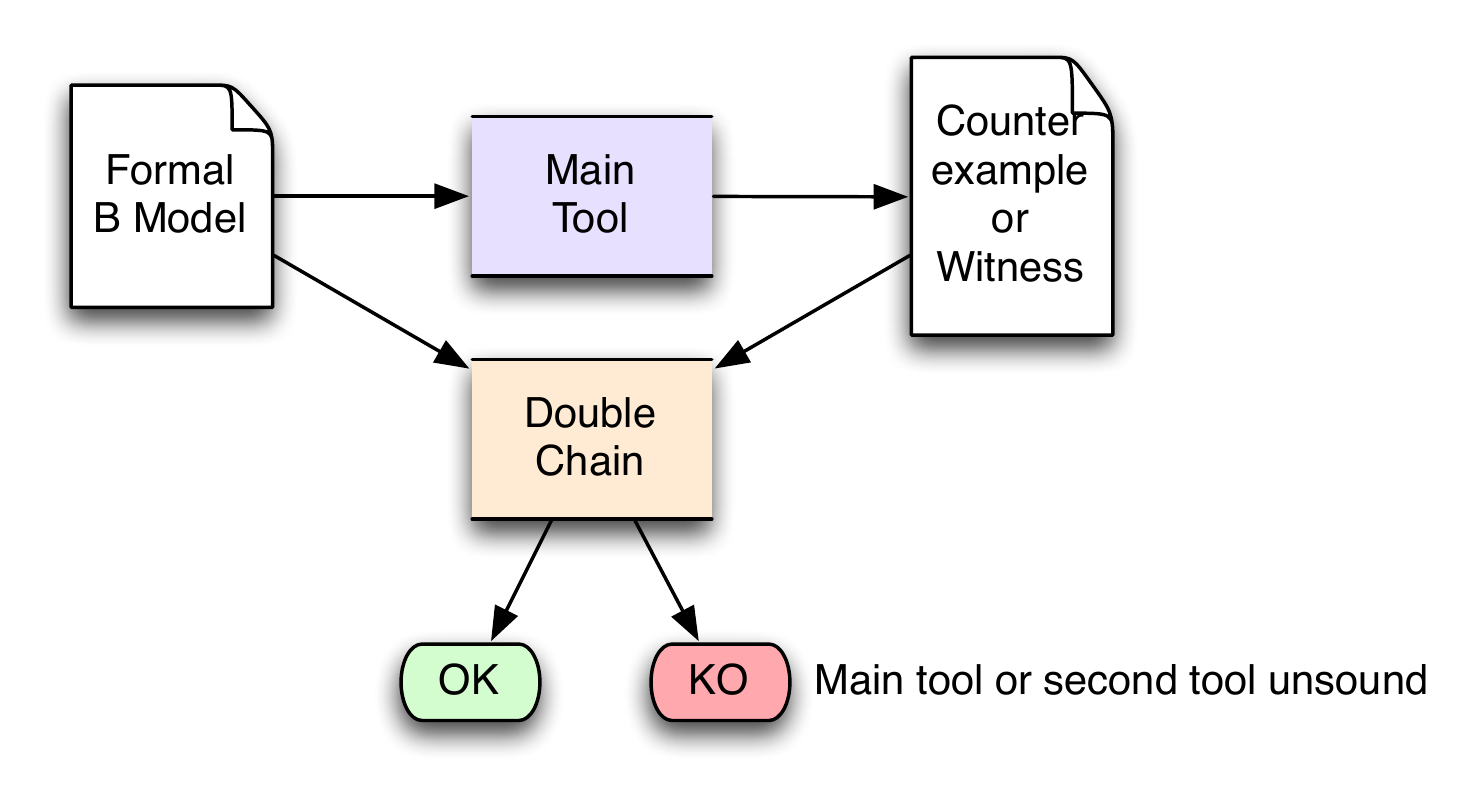}
	\caption{Double Chain}
	\label{fig:doublechain}
\end{figure}




More concretely, our long term goal is to enable \prob\ also to be used as a tool of class T3 within the norm EN 50128, i.e., moving from a tool of class T2 that 
 {\em ``supports the test or verification of the design or executable code, where errors in the tool can fail to reveal defects but cannot directly create errors in the executable software''} \cite{EN50128} to a tool that 
 {\em ``generates outputs which can directly or indirectly contribute to the executable code (including data) of the safety related system''} \cite{EN50128}.

\subsection{Approach}
The idea of a double chain is to double-check every result or output (e.g. the value of a predicate) calculated by the main tool a second time using
 an independently developed secondary toolchain (see Figure~\ref{fig:doublechain}). 
For instance, after a tool like \prob\ checks the invariant for some state of a B model, a tool like \pyb\ will 
check it again.
The main tool can provide some additional information to make the task of the secondary toolchain simpler (e.g., 
 provide witness values for existential quantifiers).
 
If the results of the two toolchains disagree, then an error is raised by the secondary toolchain and the issue will have to be investigated manually (as there
 must be a bug either in the main or in the secondary toolchain).
If the results of the two toolchains are identical, we have increased the confidence in the validity of the result.
The hope is that, in case the main tool (\prob) contains a bug, it is very unlikely that an independently developed tool
 will exhibit the same bug for the same input.
As such,
\pyb\ is a clean-room implementation which only shares the parser with \prob. 
While the kernel and interpreter of \prob\ is written in Prolog, \pyb\ is written in Python, a
dynamic imperative language.
Exactly like \prob, \pyb\ covers the B language as specified by the ClearSy B language reference manual \cite{atelierb40}.


\subsection{Outline}

In Section~\ref{main-features}, we provide a  short summery of \pyb's main features, while
 in Section~\ref{implementation} we give an overview of \pyb's implementation.
We provide our experience in developing \pyb\ in Section~\ref{developing}, which we hope will be useful
 for readers interested in embarking upon a similar path than ours.
In Section~\ref{casestudy} we present one case study and some empirical results of our tools.
We conclude with related and future work in Section~\ref{related-future}.


\section{Main Features of PyB} \label{main-features}
\prob's and \pyb's main target is the B-method. This formalism models software as abstract machines, which are written in a set based mathematical syntax and are stepwise refined to executable code. Thereby the software is seen as a state-machine in which every state must hold some safety properties.
The B formal language is based on predicate logic along with support for integers, sets, relations, functions and sequences. The syntax of B distinguishes between {\em predicates}, 
 {\em expressions\/} having a value, and {\em substitutions\/} which can modify the variables of a B machine. \\

\pyb's role is to double check \prob's results in the context of data-validation \cite{LeuschelEtAl:STS09,DBLP:journals/fac/LeuschelFFP11}.
As such, \pyb\ has to be able to:
\begin{enumerate}
\item evaluate B expressions over sets, relations, functions and sequences. As such, \pyb\ also supports set comprehensions and lambda abstractions.
\item evaluate B predicates with universal and existential quantifiers,
\item execute B substitutions (certain data validation are encoded as sequences of operations).
\end{enumerate}
\pyb\ has been integrated into \prob's Tcl-Tk version, and can be called automatically to double check the results of \prob.
Other features of the tool are a REPL (read-eval-print-loop) and its interactive animation mode. These features are
not discussed further in the present paper, as they are of limited relevance in the context of the double chain.

\section{Architecture of PyB} \label{implementation}
In this section we introduce some implementation details of \pyb. 
First, Figure~\ref{fig:modules} shows a module overview of the source code. \\
\begin{figure}[ht]
	\centering
	\caption{Module Overview}
	\label{fig:modules}
\begin{tabular}{|l|l|} \hline
\textbf{Name} & \textbf{Summery}   \\  \hline
animation\_clui.py &  console interface for animation mode  \\  \hline
animation.py & \bigcell{l}{main animation calculation, \\ together with the substitution method} \\  \hline
ast\_nodes.py &  Python classes representing AST-nodes  \\  \hline
bexceptions.py & custom exception objects   \\  \hline
bmachine.py &  a class representing one B-machine  \\  \hline
boperation.py &  a class representing one B-machine opertation  \\  \hline
bstate.py & a class representing one B-machine state  \\  \hline
btypes.py & type objects \\  \hline
config.py & main config file   \\  \hline
constrainsolver.py & B-wrapper to use a third-pary constraint solving code \\  \hline
definition\_handler.py & main definition handling code   \\  \hline
enumeration.py & enumeration methods for sets, functions, relations and more   \\  \hline
environment.py & B-state managing code   \\  \hline
external\_functions.py & implementation of external functions   \\  \hline
fake\_sets.py & implementation of large and infinite sets   \\  \hline
helpers.py &  miscellaneous helper functions  \\  \hline
interp.py & \bigcell{l}{main interpreter code. Predicate/expression evaluation code. \\ Substitution execution code}\\  \hline
parsing.py &  helper functions to execute Python AST-code  \\  \hline
pretty\_printer.py & pretty printer for b predicates and expressions   \\  \hline
pyB.py &  main module.  \\  \hline
quick\_eval.py &  helper functions to enable evaluation without enumeration  \\  \hline
repl.py &  read-eval-print-loop code  \\  \hline
statespace.py &  implementation of the state space  \\  \hline
typing.py &  main type checking code  \\  \hline
\end{tabular}
\end{figure}

\subsection{Parsing and Typing}
\pyb\ is an independent clean-room implementation, except for its Java parser.
This Java parser was written by Fabian Fritz in 2008 and is also used by \prob. 
\pyb\ uses its parser to recognise B-constructs like predicates and expressions.
These constructs are translated to an intermediate representation: an abstract syntax Tree (AST)
made of Java objects.
These Java objects are now translated to Python objects via an AST-visitor, an addition to the Java code.
This visitor is the only part of \pyb\ which is written in Java. \\

The AST-visitor emits a string of Python code. 
The dynamic features of Python enable the execution of this Python code emitted by the Java visitor.\\ 

Listing one (Figure ~\ref{fig:listingOne}) shows the Python code created by the Java-visitor
for the simple predicate $1+1=x$. 
The AST objects are numbered from 0 to 5. The last one is the root of the tree. All
these Python objects are derived from one node class. Code like this
can be evaluated by the interpreter and is the main input for most \pyb\ methods. \\

\begin{figure}[ht]
	\centering
	\caption{Python code for the predicate 1+1=x}
	\label{fig:listingOne}
\begin{small}\begin{tt}
\begin{lstlisting}[language=Python]
  id0=AIntegerExpression(1)
  id1=AIntegerExpression(1)
  id2=AAddExpression()
  id2.children.append(id0)
  id2.children.append(id1)
  id3=AIdentifierExpression("x")
  id4=AEqualPredicate()
  id4.children.append(id2)
  id4.children.append(id3)
  id5=APredicateParseUnit()
  id5.children.append(id4)
  root = id5
\end{lstlisting}
\end{tt}\end{small}
\end{figure}

Before the interpretation starts, the AST must pass a type checker, which uses a Hindley-Milner style unification algorithm. \\

The type checker processes every B expression/substitution by assigning every identifier node to a B type or type variable. Besides the primitive types integer, string, boolean and set, B introduces compound-types for cartesian products, powersets, structs and tuples, so a type variable may be a tree of concrete and unknown types.
If  some type variables (sub trees) cannot be resolved to a B type, the B file contains type errors. An example of ambiguity is the empty set or the expression \textbf{x*y}, which can be a multiplication or a cartesian product.  \\

After a successful pass of the type checking code, a type is added to every identifier node
(in this example x).
Type information is of course also important for enumeration of values, e.g., during the evaluation of quantified predicates.
The unification based algorithm makes \pyb\ compatible with 
\prob, and it is more powerful than
the type checking of other tools such as Atelier-B (because the order of the predicates is less important).
For example, the predicate {\tt x=y+1} is well typed for \pyb\ and \prob\, whereas Atelier-B requires the addition of typing information:
  {\tt\verb.x:INTEGER & y:INTEGER & x=y+1.}.\\

Between the parsing and type checking phase we also have a phase where B definitions (macros)
are substituted.
Possible external function calls --- a particular feature of \prob\ which allows linking external code with B specifications --- are also resolved at this time.

\subsection{Implementation of B's data types}
The most important data type of B is the set.
In B there exist built-in sets for boolean, natural or integer numbers. 
Relations are sets of tuples. Functions and sequences are just special cases of relations. All B-operations
like power set or the cartesian product are only producing more complex sets. \\

While booleans and integers are represented by their Python built-in counterparts, most data types are represented in \pyb\ by the Python built-in type: frozenset\footnote{http://docs.python.org/2.4/lib/types-set.html}. Frozensets are immutable 
objects which already implement all basic set operations like union, intersection, inclusion, membership etc. \\

Relations are implemented as frozensets of tuple objects, 
which are also Python built-ins. For 
example a B-function which maps the numbers 1 to 3 to its square numbers "$f=\%x.(x>0\ \&\ x<4|x*x)$" is represented on the
Python level as \textbf{frozenset([(1,1),(2,4),(3,9)])}. Objects like this can be created during the interpretation of B.\\

\subsection{Interpretation of B}
Expressions and predicates are evaluated by the interpreter module.
The main evaluation method has two parameters:
the AST root or the root of a subtree and an environment. The method call returns a value, if the tree represents 
a predicate, it returns true or false. 

The interpreter recursively performs a depth-first walk on the tree while evaluating different code for
every object class. The environment holds the state-space of the B-machine. Every state is a stack of
hash maps holding the current values of the identifiers. Every new scope (e.g. a quantified predicate) creates
a new frame on this stack. This is a standard approach of interpreter implementation. \\

\begin{figure}[ht]
	\centering
	\caption{excerpt of the interpret method (indentation of last case changed for readability) }
	\label{fig:listingTwo}
\begin{small}\begin{tt}
\begin{lstlisting}[language=Python]
    elif isinstance(node, AAddExpression):
        expr1 = interpret(node.children[0], env)
        expr2 = interpret(node.children[1], env)
        return expr1 + expr2
    elif isinstance(node, AMinusOrSetSubtractExpression) 
          or isinstance(node, ASetSubtractionExpression):
        expr1 = interpret(node.children[0], env)
        expr2 = interpret(node.children[1], env)
        return expr1 - expr2
\end{lstlisting}
\end{tt}\end{small}
\end{figure}

Figure ~\ref{fig:listingTwo} shows two of over hundred cases inside the ``interpret'' method.
The evaluation depends on the type of the 
visited AST-node. The first case is that of a simple addition. The interpreter recursively visits its subtrees
and adds the values of the calculated subexpressions. The second case is similar to the first one with only one exception:
if the expressions expr1 and expr2 are numbers a simple integer subtraction will calculated, but if they are frozensets then a set subtraction will take place. The overladed minus-operator, is natively defined on frozensets in Python. That means a method of the built-in frozenset object is executed.

\subsection{Animation of B}
B Substitutions (aka statements) are handled similar to the evaluation of predicates. The main difference are two aspects:
First, the evaluation does not return a value, but true or false if a sequence of substitutions was successfully 
executed. Second, the evaluation produces a new B state. 
This state is derived from a copy of the current state and will
be added to the state-space. Later it can become the current state (in case of interactive animation if the user
chooses this operation). \\

Sometimes a sequence of substitutions inside a B-operation consists of nondeterministic substitutions. These substitutions can be seen as choice point with different execution branches. \pyb\ explores every branch by backtracking to this choice point. A new state is returned for every possible execution branch. \\

All visited states are saved. This enables \pyb\ to backtrack on the state level. This
is an important feature needed for interactive animation and possible model checking by \pyb\ in the future.

\subsection{Difficult Aspects of B}
Checking a \prob\ solution can be very complex.
The B formulas may contain quantified predicates, set comprehensions or lambda  expressions.
The solutions computed for variables or parameters can be very large sets. \\

\pyb\ evaluates the B constructs by a brute force approach. It generates all possible values and checks
if they fulfil the constraints.
For example, \pyb\ checks an existentially quantified predicate by checking all
values of the type of the quantified variables. If no value is found, the predicate is false.\\

Of course this approach has to fail if the set of possible values is very large or infinite. Then a symbolic
representation or constraint solving can be the solution to this dilemma. \pyb\ generates special set
classes instead of frozen sets if a set becomes very large or infinite. This is not fully implemented at
the current development level. Also \pyb\ uses a external constraint solver to constrain the set of
possible solutions. This usage of constraint solving will be extended in the future too.


\subsection{Linking with ProB}
\subsubsection{Verification of States}
\begin{figure}[ht]
	\centering
  	\includegraphics[width=300px]{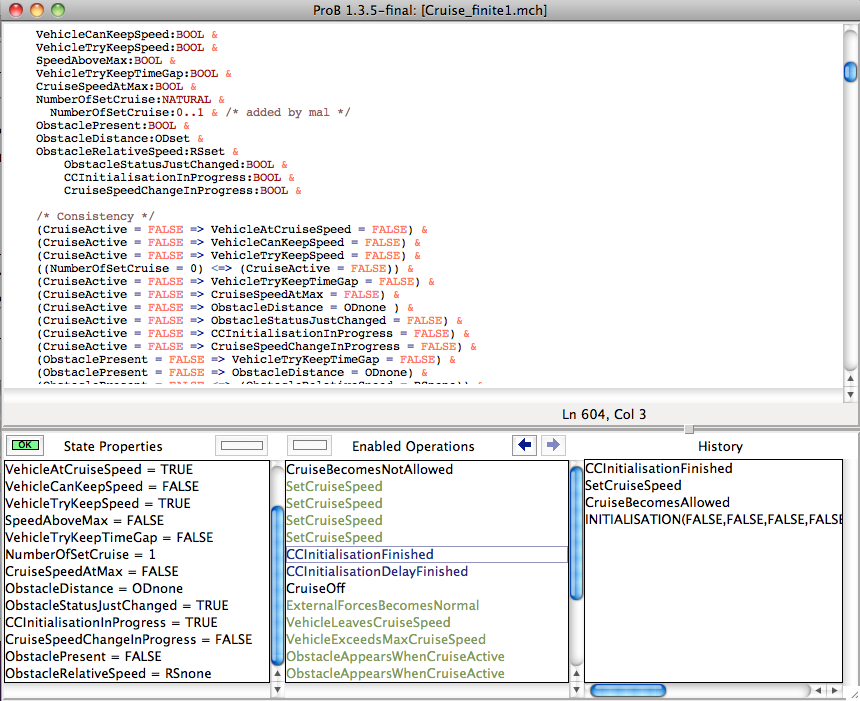}
	\caption{\textbf{ProB} animating a B-Machine: The B-Machine code, the B-State (values and constants), enabled operations, history}
	\label{fig:prob}
\end{figure}

Below we use an example of a complex B-Machine (cruise control model) with a simple state.
Figure ~\ref{fig:prob} shows \prob\ in its animation-mode for this example.
At some point the user can save the B-State to a file (Figure ~\ref{fig:state} ). \\
This solution-file contains a list of constant- and variable values. The right side of each equation
may be any B-formula: a number, a set, a relation, a function or even a lambda-expression. \\
Eventually \pyb\ reads this file, generates a B-State, evaluates the Properties- and Invariantclause of the B-file and outputs if a safety property was violated in this state.
Currently this process is automated via a Python script but will be fully included into the official \prob\ release in the future.\\

\begin{figure}[ht]
	\centering
	\begin{small}\begin{tt}
\begin{lstlisting}
 /* Variables */
  #PREDICATE
    CruiseAllowed = FALSE
  & CruiseActive = FALSE
  & VehicleAtCruiseSpeed = FALSE
  & VehicleCanKeepSpeed = FALSE
  & VehicleTryKeepSpeed = FALSE
  & SpeedAboveMax = FALSE
  & VehicleTryKeepTimeGap = FALSE
  & NumberOfSetCruise = 0
  & CruiseSpeedAtMax = FALSE
  & ObstacleDistance = ODnone
  & ObstacleStatusJustChanged = FALSE
  & CCInitialisationInProgress = FALSE
  & CruiseSpeedChangeInProgress = FALSE
  & ObstaclePresent = TRUE
  & ObstacleRelativeSpeed = RSequal
\end{lstlisting} \end{tt}\end{small}
	\caption{A simple input example of \textbf{PyB}: This file contains all values and constants of a B-State, The first line \#PREDICATE was added for parsing reasons}
	\label{fig:state}
\end{figure}

\begin{figure}[ht]
	\centering
  	\includegraphics[width=400px]{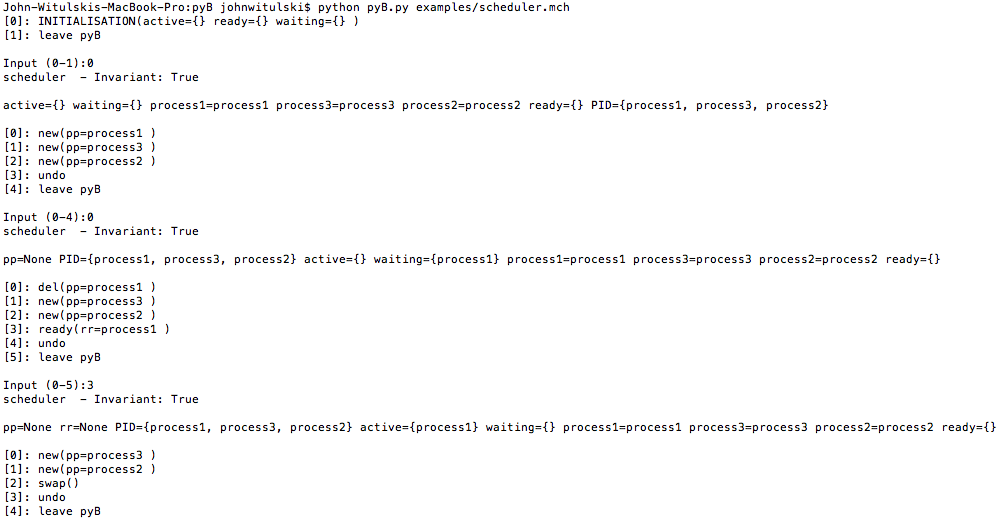}
	\caption{Console output of PyB while animation a B-model. PyB prints the current state, the status of the invariant and the enabeld operations}
	\label{fig:anim}
\end{figure}




\subsubsection{Verification of State-changes}
Another application of \pyb\ as second toolchain is not the verification of states, but the execution of operations 
(i.e., the application of substitutions). In this approach, it is not of interest if a computed state satisfies a safety
condition, but rather which operations are enabled at a specific state and if the 'execution' of an operation 
leads to the same state computed by other tools like \prob\. \\

In this case, the second toolchain is not used to check if a safe or faulty state is really safe or faulty, 
but if the states (that violate the safety properties of a model) are really reachable. \\

Figure ~\ref{fig:anim} shows the console output of an interactive animation of a simple scheduler B model.



\section{Development Experience of PyB} \label{developing}

\subsection{Timeline and Effort}
The developmet of \pyb\ started late 2011 and continued until today. 
At approximately 10 hours
work per week on average, the current tool is a result of approximately 1,000 hours of work. It consists of
over 7400 lines of code and 12000 lines of test-code. \

Difficult parts of the implementation where type checking and the execution of nondeterministic
substitutions. Especially the type checking implementation via unification consumed some time to assure
compatibility with \prob. As usual small bugs caused by missing specification details inside the implementations 
consumed a lot of time. \\

Development was done using a version-control system. The progress of the project can be tracked 
via the commit-messages\footnote{https://github.com/hhu-stups/pyB/commits/} of the git repository. 
Figure ~\ref{fig:timeline} shows the project timeline.\\ 

\begin{figure}[ht]
	\centering
	\caption{Project timeline}
	\label{fig:timeline}
 \begin{tabular}{|l|l|} \hline
    \textbf{Date} & \textbf{milestone}   \\  \hline
    August 2011 & project start   \\  \hline
    September 2011 & evaluation of simple arithmetic, set-predicates, functions and relations   \\  \hline
    October 2011 & type-checking of simple arithmetic, set-predicates, functions and relations   \\  \hline
    November 2011 & \bigcell{l}{type-checking with simple unifications and \\ replaced interpreter state by an more complex environment}  \\  \hline
    December 2011 & \bigcell{l}{typing and evaluation of more complex constructs. \\ Added a simple (brute force) enumerator. First parsing of whole B-machines}  \\  \hline
    January 2012 & first evaluation of simple B-machine assertions   \\  \hline
    February 2012 & \bigcell{l}{implementation of more complex functions like closure and UNION. \\ First evaluation of simple B-machine PROPERTIES-, \\ CONSTANT- and DEFINITION-clauses} \\  \hline
    March 2012 & implementation of IF-THEN, CHOICE and SELECT-substitutions   \\  \hline
    April 2012 & implementation of lookup of SEEN or INCLUDE B-machines  \\  \hline
    September 2012 & implemented quick eval functions to speed up tool performance  \\  \hline
    October 2012 & first successful usage of an extern constraint solver  \\  \hline
    November 2012 & first introduction of state-space.   \\  \hline
    December 2012 & successful animation of simple B-machines.  \\  \hline 
    January 2013 & implementation of a small B-REPL  \\  \hline
    February 2013 & successful usage of ProB solutions  \\  \hline
    March 2013 & successful run of alstom case-study  \\  \hline
    April 2013 & complex animation-refactoring to enable nondeterministic substitutions  \\  \hline
    May 2013 & animation of SEEN or INCLUDED B-machines/operations  \\  \hline
    June 2013 & documentation of tool-features and implementation details  \\  \hline
    July 2013 & implementation of complex nondeterministic substitutions  \\  \hline
    August 2013 & \bigcell{l}{implementation of nondeterministic set\_up\_constants and init phase. \\ Added pretty printer for predicates.}  \\  \hline
    September 2013 & typing and execution of external functions  \\  \hline
    October 2013 & usage of more complex ProB solutions \\  \hline
    November 2013 & added symbolic representation for large and infinite sets \\  \hline
    December 2013 & some systerel (industrial B-machines) successfully checked with pyB/ProB \\  \hline
 \end{tabular}
\end{figure}

\subsection{Testing and Validation}
\pyb\ has been developed using test-driven development (TDD). The process of TDD is as follows:
\begin{enumerate}
 \item Test-code is written for an unimplemented new feature. This code is called test-case.
 \item The first run (execution) of the test-case fails. I.e all assertions inside this test-case are false. 
 This prevents the programmer from implementing an already implemented feature.
 \item The feature is implemented and the test passes: All assertions of the test-case are true.
 \item The code is refactored. 
 \item All previous written test-cases also pass. This ensures that the new implementation has not
 destroyed any previous functionality.
\end{enumerate}
This process is automated to some degree. Because the tool was implemented by only one programmer,
the distributed aspects and advantages of TDD are omitted in this overview.\
In early development-stages ASTs where constructed explicit by creating tree-nodes as input to 
the interpreter. Examples of assertions where simple arithmetic or boolean properties. When the 
development of the tool proceeds, the test-case become more complex.\

After the successful usage of the Java parser, ASTs were created automatically by the parsing-module.
Inputs also become more complex. Starting with easy inputs like `x=y+42`, the tests quickly also includes sets, functions and relations. All this easy tests are still present and passed by \pyb \

At the current development new test-cases are full B-machines. Assertions 
are not longer made just on single predicates, but made up of whole B-machine states.
Examples of assertions are true or false properties/ assertions/ invariants after the B-machine initialization,
enabled or disabled operations inside a specific state or after some animation steps and of course
the test if \pyb\ gets the same result as \prob. \

Of course TDD can not guarantee the same reliability than a formal proof of \pyb.
But it still makes \pyb\ more reliable and it is a much better approach than testing the tool afterwards: TDD reveals errors inside an implementation very soon. \

\section{Case Studies and Empirical Evaluation} \label{casestudy}

\subsection{Alstom Case Study}
The case study consists of 6 industrial B-machines provided by Alstom. Every machine was model-checked with \prob. 
Two machines where faulty. They defined a partial surjection to an infinite set and initialized it with the empty set. 
After the correction to a partial function the machine still contains a deadlocked state. 

The procedure of the double-checking was performed as follows:\\
Every machine was animated n-times with \prob. After every animation-step the state of 
the machine (only constants and variables) where written to a file. The data was loaded by \pyb. 
\pyb\ evaluated the properties and invariant of the machine and returned the result. 
	
After the configuration of some tool properties (maximum size of integer-sets), \pyb\ successfully checked 3 of 6 B-machines
by double checking of 32 to 42 states of the machines in 5 minutes per machine. One B machine doesn't work with \pyb\ 
because of the missing support of external functions like append (on B strings). The remaining machines fail at the same 
point as \prob\ (described above). The animation with \pyb\ alone (without \prob's information) of all machines fails.


\subsection{Performance Evaluation}
\pyb\ is still in an early development phase. It may be possible to speed up checking by 
replacing data transfer by a socket communication (about 40\% of the runtime) with \prob\ or refactoring the tool
using object dispatching instead of the expensive Python ``isinstance'' built-in.
Also using the pypy just-in-time compiler technology on \pyb\ seams promising \\

Also the generation of large sets induces a serious performance issue.
The evaluation of a cartesian product of two sets or the calculation of the power set of a set of 19 elements needs more than 7 seconds.
The calculation of the power set of 22 elements already needs more than 440 seconds.
Some performance issues can only be solved by better constraint solving. \\

\section{Related and Future Work} \label{related-future}
\subsection{Related Work}\label{relatedWork}
PredicateB and PredicateB++ are similar tools to pyB. In contrast to pyB they only evaluate predicates and 
need additional software like ovado\footnote{More Informations about Ovado and PredicateB at http://www.data-validation.fr/}\cite{DBLP:journals/corr/abs-1210-6815} 
or the DVT tool to validate data like B-states.
Predicate B (written in Java) and PredicateB++ (in C++) were 
created by the company ClearSy in 2005 and 2008.
They have been successfully used by Brama \cite{Brama:B2007} inside the 
Rodin Platform \cite{rodinplatform}.
Another, recent tool is the JEB animator \cite{DBLP:conf/apsec/YangJS12}  written in JavaScript.
Systerel + Ovado \cite{DBLP:journals/corr/abs-1210-7039}

\nocite{DBLP:journals/sttt/LeuschelB08} 
\nocite{LeuschelEtAl:STS09,DBLP:journals/fac/LeuschelFFP11} 


Outside of the B community, there is of course considerable work on proving tools correct.
We just want to mention the grand challenge of the verifying compiler \cite{hoare03} and the
 work on the Compcert verified compiler (e.g., \cite{Tristan-Leroy-scheduling}).

\subsection{Future Work}\label{Future Work}
The most important issue that has to be solved, to guarantee the correct checking of solutions generated by other tools (like \prob), is the handling of large or infinite sets and the full usage of a better external constraint-solver.
Even if all variables and constants of a state are known (calculated by an other tool), there may be (quantified)
predicates for which checking could cause the generation of large sets.

At the current development state there are critical performance issues. Some of them will be solved by a more 
efficient implementation of the tool. Other problems are a result of the choice of Python as implementation language. 
These problem can be solved by refactoring the tool to R-Python, a statically typeable subset of Python which can be translated to c using the pypy technology\footnote{Using pypy is one reason for choosing the Python language}. Performance can also be increased by the use of a better constraint-solver.\\



A completely different aspect of this research is to see how fast can this tool be by using the pypy just-in-time compiler technology. 
The tool is not far away from becoming a full model-checker when a constraint-solver is successfully integrated 
in this project. Also the tool is able to animate B-machines. This will be useful to check if the animation of 
\prob\ is correct, i.e if the same states are enabled and the same deadlocks are found.



\subsection{Conclusion}\label{conclusion}
\pyb\,  has been successfully used to validate data generated by \prob\ for many simple B machines and some more complex, industrial B 
machines.
Once all performance issues are solved, we will have a reliable, independently developed, double chain
 for model checking and data validation for B models.


\bibliographystyle{eptcs}

\bibliography{michael}


\end{document}